\documentclass[nohyper,12pt,letterpaper]{JHEP3}
\usepackage{youngtab,epsfig}

\author{Robert de Mello Koch and Rhiannon Gwyn\\
\qquad \\
Department of Physics and Centre for Theoretical Physics,\\ 
University of the Witwatersrand,\\ 
Wits, 2050,\\ 
South Africa\\
\qquad\\
Stellenbosch Institute for Advanced Studies,\\
Stellenbosch,\\
South Africa\\
\qquad\\
E-mail: \email{robert,rhiannon@neo.phys.wits.ac.za}}

\abstract{
Certain correlation functions are computed exactly in the zero coupling
limit of ${\cal N}=4$ super Yang-Mills theory with gauge group $SU(N)$.
A set of linearly independent operators that are in one-to-one 
correspondence with the half-BPS representations of the $SU(N)$ gauge 
theory is given. These results are used to study giant gravitons in the 
dual $AdS_5\times S^5$ string theory. In addition, for the $U(N)$ gauge 
theory, we explain how to systematically identify contributions coming from 
the boundary degrees of freedom.}

\preprint{}

\title{Giant Graviton Correlators from Dual $SU(N)$ super Yang-Mills Theory}


\def \Tr{\mbox{Tr\,}}

\begin{document}

\section{Introduction}

Non-perturbative string theory can now be studied, thanks to the AdS/CFT 
correspondence, using techniques in the dual conformal field 
theory\cite{AdSCFT}. It was in this spirit that giant graviton correlators
were studied in \cite{JeRa},\cite{CoRa}. By studying the zero coupling limit of
${\cal N}=4$ super Yang-Mills theory with gauge group $U(N)$, candidate 
operators dual to giant gravitons were proposed. Further, a powerful 
machinery allowing the exact computation of a class of correlators at 
finite $N$ was developed. 

In this article we take the first steps towards providing a dictionary
between half-BPS representations in the ${\cal N}=4$ super Yang-Mills theory 
with gauge group $SU(N)$ and states in the dual supergravity. For this purpose,
it is enough to study the zero coupling limit of the Yang-Mills theory. Our goal
was to provide a one-to-one correspondence between the half-BPS representations 
and a set of operators that have diagonal two point functions. These operators
(or ones related to them by a unitary transformation acting on the space of
normalized operators and hence preserving the two point functions) would then 
be natural candidates for particle states in the supergravity. We have been 
partially successful.

We will now explain our interest in the extension to gauge group $SU(N)$. A $U(N)$
gauge theory is equivalent to a free $U(1)$ vector multiplet times an $SU(N)$ gauge 
theory up to some $Z_N$ identifications. The $U(1)$ vector multiplet is related to
the centre of mass motion of all the branes\cite{Gibbons}. These modes live at the 
boundary and are called singletons or doubletons\cite{singletons}. Therefore, the 
bulk $AdS$ theory (in which we are interested) is describing the $SU(N)$ part of the 
gauge theory\cite{review}. One of the things which interests us is the extent to
which the bulk and boundary degrees of freedom can be separated in the dual super
Yang-Mills theory. 

One may have expected that the generalization from gauge group $U(N)$ to $SU(N)$ 
would imply relatively minor modification of the results of \cite{JeRa}. This is 
not the case as we now explain. In the $U(N)$ case, the space of Schur polynomials 
can be mapped to the space of half-BPS representations. Further, the Schur 
polynomials diagonalise the two point function. In the $SU(N)$ case, the number
of Schur polynomials is not even equal to the number of half-BPS representations.
The Schur polynomials are no longer orthogonal - in fact, they are not even
linearly independent. Despite this, using the results of \cite{JeRa}, we show that
they are still a useful set of operators to consider. Indeed, by studying Schur
polynomials, we are able to generalize many of the results of \cite{JeRa} to the 
$SU(N)$ case. We are also able to select a set of linearly independent operators 
that are in one-to-one correspondence with the half-BPS representations of the 
$SU(N)$ gauge theory.

Our article is organized as follows: In section 2 we give a brief review of the relevant
results from \cite{JeRa}. In section 3 we develop the technology needed to study correlation
functions in the $SU(N)$ gauge theory. In section 4 we use these results to study giant
gravitons in the $AdS_5\times S^5$ string theory.

\section{Review of $U(N)$ Technology}

Corley, Jevicki and Ramgoolam have developed a powerful machinery for the 
exact computation of a class of correlators in the zero coupling limit of 
${\cal N}=4$ super Yang-Mills theory with gauge group $U(N)$\cite{JeRa}. The 
operators considered in \cite{JeRa} are half-BPS chiral primary operators built 
from a single complex combination (denoted $\Phi$ in what follows) of any two 
of the six Higgs fields appearing in the theory. Using a total of $n$ $\Phi$s,
there is a distinct operator for each partition of $n$. Further, there is
a one-to-one correspondence between these operators and half-BPS 
representations of ${\cal R}$ charge $n$. The Schur Polynomials of degree $n$
provide a useful basis in this space of operators\cite{JeRa}. Indeed, in the
next section we will summarize exact results obtained in \cite{JeRa} for the 
correlation functions of Schur Polynomials. We then recall the application
of these results to the physics of giant gravitons. Finally, we review
Berenstein's argument \cite{Berenstein} which essentially explains why Schur 
polynomials behave as D-branes.

\subsection{Exact Correlators}

All correlators are computed using the free field contraction

\begin{equation}
\langle\Phi_{ij}(x)\Phi^*_{kl}(y)\rangle =
{\delta_{ik}\delta_{jl}\over (x-y)^2}.
\label{contraction}
\end{equation}

\noindent
Consider a representation $R$ labelled by a specific Young diagram containing $n_R$ boxes. 
This Young diagram labels both a representation of $U(N)$ and a representation of
$S_{n_{R}}$. Schur polynomials

$$\chi_R (\Phi )={1\over n!}\sum_{\sigma\in S_n}\chi_R(\sigma )\Tr (\sigma\Phi ),$$

\noindent
where

$$\Tr(\sigma\Phi )\equiv \sum_{i_1, i_2,\cdots i_n}\Phi^{i_1}_{i_{\sigma(1)}}
\Phi^{i_2}_{i_{\sigma(2)}}\cdots \Phi^{i_n}_{i_{\sigma(n)}},$$

\noindent
have a particularly simple two point correlation function 

\begin{equation}
{\langle\chi_R (\Phi (x))\chi_S(\Phi^* (y))\rangle =\delta_{RS}{D_R n_R!\over d_R}
{1\over (x-y)^{n_R}}.}
\label{TwPnt}
\end{equation}

\noindent
In this last formula, $D_R$ is the dimension of the representation $R$ of the unitary group
and $d_R$ is the dimension of representation $R$ of the permutation group. The fact that
the two point function is diagonal is significant as explained below. The spacetime 
dependence in (\ref{TwPnt}) is trivial. The non-trivial part of the result is contained in 
the factor obtained from the sum over $U(N)$ indices. For this reason, from now on, we 
suppress the spacetime dependence in all formulas. There is an equally elegant 
result for three point correlators: they are directly related to fusion coefficients.
For a derivation of these results and further results for higher point functions, the 
interested reader is referred to \cite{JeRa}.

A comment is in order: the Schur polynomials are the characters of the unitary group
in their irreducible representations. The complex matrix $\Phi =\phi_1+i\phi_2$ with
$\phi_1$ and $\phi_2$ $u(N)$ valued. Thus, we are considering an extension of the Schur
polynomials from unitary matrices to complex matrices. These polynomials form a basis 
for the $U(N)$ invariant functions of $\Phi$, with $\Phi$ transforming in the adjoint
representation.

\subsection{Giant Gravitons}

Traces involving $n<\!\!< N$ $\Phi$ fields do not mix in the large $N$ limit. Indeed,
normalizing our operators so that the leading contribution to the two point function
is independent of $N$ we have

$$\langle {\Tr (\Phi^{n})\over N^{n\over 2}}
{\Tr (\Phi^{*n})\over N^{n\over 2}}\rangle=n(1+O(N^{-2})).$$

\noindent
With this normalization

$$ \langle {\Tr (\Phi^{n_1})\over N^{n_1\over 2}}
{\Tr (\Phi^{n_2-n_1})\over N^{n_2-n_1 \over 2}}
{\Tr (\Phi^{*n_2})\over N^{n_2\over 2}}\rangle =0+O({1\over N}).$$

\noindent
These operators correspond to Kaluza-Klein states via the AdS/CFT correspondence.
Multi-trace operators would correspond to multi-particle states, with the number 
of traces in the gauge theory operator matching the number of particles in the 
supergravity state. As $n$ increases, mixing between operators is no longer suppressed
and the dictionary between single traces and single particle states is modified.
To correct the dictionary, one needs to find linear combinations of operators
that again do not mix. These are the operators that will have a particle interpretation
in the supergravity. Since the two point function of the Schur polynomials is diagonal, 
they provide a natural candidate.

To provide a more detailed interpretation for the Schur polynomials, recall that in background 
fields, branes can get polarized into higher branes\cite{Myers}. Using this insight, giant
gravitons were discovered in \cite{Lenny} as solutions to the equations of motion following
from brane actions. The giant graviton solutions describe branes extended in the sphere 
of the $AdS\times S$ background. These are the so-called sphere giants. The larger the 
angular momentum of the graviton, the larger the sphere giant. The size of these branes 
is limited by the radius of the sphere, thereby providing a natural cut-off on the 
angular momentum of gravitons in this background. In addition to these sphere giants, 
giant gravitons extended in the AdS space, AdS giants, were also discovered\cite{AdS}. 
In contrast to the sphere giants, the angular momentum of the AdS giants is not cut off.

Using techniques in the dual conformal field theory, finite $N$ truncations in BPS
spectra were studied as evidence of a stringy exclusion principle\cite{SEP}. This
stringy exclusion principle can be interpreted as evidence for non-commutative
gravity \cite{NCG}. The above cut-off on the angular momentum of sphere giants
seems to provide a simple explanation of the stringy exclusion principle. However,
the fact that the angular momentum of AdS giants is not cut off obscures this 
connection. Giant gravitons probably do explain the stringy exclusion principle, since 
there is some evidence that AdS giants with angular momentum exceeding the stringy
exclusion principle bound are not BPS\cite{Julian}.

The number of boxes in the totally antisymmetric representation of $U(N)$ is cut off
at $N$, whilst the number of boxes in the totally symmetric representation is not cut 
off. This has a natural physical interpretation as we now explain. The number of
boxes in the Young diagram is equal to the degree of the corresponding Schur polynomial,
and hence to the ${\cal R}$ charge of the operator. According to the AdS/CFT correspondence,
the ${\cal R}$ charge of the super Yang-Mills operator maps into the angular momentum 
of the dual supergravity state.
This naturally suggests\cite{JeRa} that the Schur polynomial for the totally 
antisymmetric representation is dual to a sphere giant, whilst the Schur polynomial
for the totally symmetric representation is dual to an AdS giant\footnote{See also 
\cite{Vijay}.}. Convincing support for this conjecture is that these states have a 
well defined $1/N$ expansion and can accommodate a spectrum of open strings\cite{Aharony}.
A Schur polynomial corresponding to a more general representation would correspond
to a composite state involving giants and Kaluza-Klein gravitons.

\subsection{The Schur Polynomial/D-brane Correspondence}

In \cite{Berenstein} the large $N$ gauged quantum mechanics for a single Hermitian
matrix with quadratic potential was connected with a decoupling limit of ${\cal N}=4$
super Yang-Mills theory. This correspondence is very similar to the one exhibited in
\cite{BMN}. Three (a priori) distinct descriptions of the spectrum of the model were given.
The first description employing single trace operators is naturally related to closed
string states in the dual gravitational theory. The second description 
involves integrating the angular degrees of freedom out, so that a description in terms 
of eigenvalues is obtained. The integration introduces the square of the Van der Monde 
determinant, which is conveniently accounted for by a similarity transform after which
the eigenvalues become $N$ free fermions in the harmonic oscillator potential\cite{BIPZ}.
Recently the proposal of \cite{Berenstein} was used to provide a beautiful map between 
states of the Fermi theory and IIB supergravity geometries\cite{Lin}.
The third description of the spectrum, using the results of \cite{JeRa}, employs a
Schur polynomial basis. A surprising result\footnote{This connection was anticipated in
\cite{Andre}. See also \cite{Antal} where it is shown that the exact eigenstates of cubic
collective field theory are given by the $N$-fermion wave functions or by the Schur 
polynomials.} of \cite{Berenstein} is that the eigenvalue and Schur polynomial descriptions
coincide!

This correspondence between the Schur polynomial description of the spectrum and the
eigenvalue description essentially explains the correspondence between
sphere giants and Schur polynomials corresponding to the totally antisymmetric representation
and AdS giants and Schur polynomials corresponding to the totally symmetric representation.
Using the map between Schur polynomials and the free fermion (i.e. eigenvalue) 
descriptions\cite{Berenstein},\cite{ff} we know that the totally symmetric representation
corresponds to taking the top most eigenvalue of the Fermi sea and giving it a large energy
$n$ and that the totally symmetric representation corresponds to creating a hole deep in
the Fermi sea. In the $c=1$ matrix model\cite{mm}, these single eigenvalue excitations 
correspond to D-brane states. The above correspondence makes it clear that the giant
graviton operators proposed in \cite{JeRa} are also describing the dynamics of a single
eigenvalue, thereby explaining why these particular Schur polynomials are dual to 
D-branes\footnote{This does not prove the correspondence between Schur polynomials and
giant gravitons. Rather, this argument relates the conjectured equivalence of branes
and fermions in the $c=1$ matrix model (which is on a firm footing) with the correspondence
between Young diagrams and giant gravitons in the AdS/CFT setting.}.

\section{Technology for $SU(N)$}

In this section we use the results of \cite{JeRa} to develop techniques which allow an
efficient computation of correlation functions of Schur polynomials $\chi_R(\Psi )$
with $\Psi=\phi_1 +i\phi_2$ for any two ($\phi_1,\phi_2 \in su(N)$) of the six Higgs 
fields appearing in the theory. We will also give an algorithm which allows the 
construction of a complete basis in the space of gauge invariant operators constructed 
by taking traces of the $\Psi$s. This represents the required generalization of the 
results obtained in \cite{JeRa}. For complementary results relevant for the $SU(N)$ 
case, the reader is referred to section 10 of \cite{CoRa}. The method 
used in our work is a completely different approach. Where our results overlap, we
have checked that they agree. To simplify the notation in what
follows we use $\Phi=\phi_1+i\phi_2$ if $\phi_1,\phi_2$ are $u(N)$ valued and
$\Psi=\phi_1+i\phi_2$ if $\phi_1,\phi_2$ are $su(N)$ valued.

\subsection{A First Look}

The essential difference between the situation considered in \cite{JeRa} 
and the situation considered in this work, is that our $\Psi$ field
is traceless. This has far reaching consequences. To illustrate this 
point, consider the Schur polynomials built using two $\Psi$s. There 
are two possible representations: $\yng(1,1)$ and $\yng(2)$\,\,. The 
Schur polynomials corresponding to these two representations are given by

$$\chi_{\yng(1,1)}={1\over 2}\left( (\Tr\Psi )^2-\Tr (\Psi^2)\right),\qquad 
\chi_{\yng(2)}={1\over 2}\left( (\Tr\Psi )^2+\Tr (\Psi^2)\right).$$

\noindent
Evaluating these Schur polynomials on the (traceless) $\Psi$ field gives

$$\chi_{\yng(1,1)}= -\chi_{\yng(2)}\,\,.$$

\noindent
This example clearly demonstrates that Schur polynomials corresponding to 
different representations are no longer orthogonal (or even linearly independent).
The Schur polynomials are not in a one-to-one correspondence with the space of
half-BPS representations and the two point function is no longer diagonal. In the
remainder of this section we will argue that it is still useful to consider
correlation functions of the $\chi_R(\Psi )$.

\subsection{Recycling $U(N)$ Results}

The contraction (\ref{contraction}) for two $\Phi$ fields is to be replaced by
(spacetime dependence suppressed)

\begin{equation}
\langle\Psi_{ij}(x)\Psi^*_{kl}(y)\rangle =
\delta_{ik}\delta_{jl}-{1 \over N} \delta_{ij}\delta_{kl}.
\label{newcontraction}
\end{equation}

\noindent
One way to interpret the above formula, is that the second term implements the 
tracelessness of $\Psi$ by subtracting the contribution coming from the mode associated
with the trace. We could also perform this subtraction with the help of a ghost field $c$
with contractions

$$\langle cc\rangle =0=\langle c^* c^*\rangle,\qquad \langle cc^*\rangle =-{1\over N}=
\langle c^* c\rangle .$$

Concretely, we have the identity

$$\langle F(\Psi_{ij},\Psi^*_{kl})\rangle =
\langle F(\Phi_{ij}+c\delta_{ij},\Phi^*_{kl}+c^*\delta_{kl})\rangle ,$$

\noindent
where $F(\cdot,\cdot)$ is an arbitrary function. The advantage of trading $\Psi$ 
for $\Phi+c$ follows as a consequence of a particularly simple expansion of the 
Schur polynomial $\chi_R(\Phi+c)$ as a series in $c$. The coefficients in this 
expansion are themselves Schur polynomials in $\Phi$ so that the results of 
\cite{JeRa} can again be used, providing an efficient computational tool.

Consider a representation $R$ with $n_R$ boxes. The expansion we wish to develop 
takes the form

$$\chi_R (\Psi_{ij})\to\chi_R(\Phi_{ij}+\delta_{ij}c)
=\sum_{m=0}^{n_R}{c^m\over m!}D^m\chi_R (\Phi_{ij}),\qquad D\equiv
\sum_{i=1}^N{\partial\over \partial\Phi_{ii}}.$$

\noindent
Recall that to each box in a Young diagram we can associate a weight \cite{georgi}.
In terms of these weights there is a particularly simple expression for the action of
$D$ on the Schur polynomial

$$ D\chi_R (\Phi )=\sum_{S}f_S\chi_S (\Phi ),$$

\noindent
where the sum runs over all Young diagrams $S$ that can be obtained by removing a
single box from $R$ to leave a valid Young diagram
and the coefficient $f_S$ is the weight of the removed box. It
is straightforward to verify this rule using the explicit expressions for the 
Schur polynomials. As an illustration of the rule, consider the example

$$ D\chi_{\yng(3,2,1)}(\Phi)=(N+2)\chi_{\yng(2,2,1)}(\Phi)
+N\chi_{\yng(3,1,1)}(\Phi)+(N-2)\chi_{\yng(3,2)}(\Phi).$$

\noindent
The action of higher powers of $D$ is obtained by iterating this action, which
provides us with all the tools we need for the expansion of 
$\chi_R (\Phi_{ij}+c\delta_{ij})$. 

These results allow us to efficiently compute the correlation functions of an
arbitrary $n$ point function of the Schur polynomials $\chi_S(\Psi)$. To illustrate 
this point consider the computation of

$$\langle \chi_{\yng(1,1,1)}(\Psi )\chi_{\yng(2,1)}(\Psi^* )\rangle =
\langle \chi_{\yng(1,1,1)}(\Phi +{\bf 1}c)\chi_{\yng(2,1)}
(\Phi^* +{\bf 1}c^*)\rangle .$$

\noindent
Using the expansions

$$\chi_{\yng(1,1,1)}(\Phi +{\bf 1}c)=\chi_{\yng(1,1,1)}(\Phi )+(N-2)c
\chi_{\yng(1,1)}(\Phi)+(N-2)(N-1){c^2\over 2!}\chi_{\yng(1)}(\Phi )+
(N-2)(N-1)N{c^3\over 3!},$$

\begin{eqnarray}
\chi_{\yng(2,1)}(\Phi^* +{\bf 1}c^*)=&&
\chi_{\yng(2,1)}(\Phi^*)+
(N+1)c^*\chi_{\yng(1,1)}(\Phi^*)+
(N-1)c^*\chi_{\yng(2)}(\Phi^*)\nonumber\\
&+&2(N^2-1){c^{*2}\over 2!}\chi_{\yng(1)}(\Phi^*)+
2N(N^2-1){c^{*3}\over 3!},
\nonumber
\end{eqnarray}

\noindent
and the formula

$$\langle c^n c^{*m}\rangle ={\delta^{mn}(-1)^n n!\over N^n} ,$$

\noindent
we easily obtain

\begin{eqnarray}
\langle \chi_{\yng(1,1,1)}(\Psi )\chi_{\yng(2,1)}
(\Psi^* )\rangle =&&
\langle \chi_{\yng(1,1,1)}(\Phi )\chi_{\yng(2,1)}(\Phi^* )\rangle
\nonumber \\ 
&-&{1\over N}\langle (N-2)\chi_{\yng(1,1)}(\Phi)
\left[ (N+1)\chi_{\yng(1,1)}(\Phi^*)+
(N-1)\chi_{\yng(2)}(\Phi^*)\right]\rangle
\nonumber \\
&+&{1\over N^2}\langle (N-2)(N-1)\chi_{\yng(1)}(\Phi)
\left[ (N^2 -1)\chi_{\yng(1)}(\Phi^*)\right]\rangle
\nonumber \\
&-&{2N^2(N^2-1)(N-2)(N-1)\over 3!N^3}
\nonumber
\end{eqnarray}

\noindent
The remaining correlators can now all be evaluated using (\ref{TwPnt}). The results of this
section could also have been obtained without introducing the ghost $c$ and simply using
the contraction (\ref{newcontraction}). This is the approach taken in \cite{CoRa}. We have
checked that our results are in complete agreement.
The generalization to $n$ point functions is obvious.

\subsection{Constraints}

We have already seen that because $\Tr (\Psi )=0$, not all Schur polynomials 
$\chi_R(\Psi )$ are linearly independent. In this section we obtain a complete
set of linear relations between the $\chi_R(\Psi )$s. For a different use of
the $\Tr (\Psi )=0$ condition, consult\cite{CoRa}.

Recall that the character for a group element $T$ in a direct product representation 
$\chi_{R\times S}(T)$ is equal to the product of the characters $\chi_R(T)\chi_S(T)$.
The characters of the unitary group are given by the Schur polynomials. Thus,
the Schur polynomials themselves obey these relations. The fact that we evaluate the
Schur polynomials on the complex matrix $\Psi$ and not on a unitary matrix is of no
consequence. Using this insight, we can write down expressions obtained by multiplying
$\yng(1)$ with an arbitrary representation $R$. Upon noting that

$$\chi_{\yng(1)}(\Psi )= \Tr(\Psi) =0,$$

\noindent
we see that this leads to a set of constraints obeyed by the $\chi_R(\Psi )$. As an example

$$\chi_{\yng(1)}(\Psi )\chi_{\yng(1,1)}(\Psi )= 
\chi_{\yng(1,1,1)}(\Psi )+
\chi_{\yng(2,1)}(\Psi )=0$$

\noindent
which is easily verified explicitly. The number of relations between the Schur
polynomials corresponding to Young diagrams with $n$ boxes obtained using the
above procedure is equal to the number of Schur polynomials corresponding to
Young diagrams with $n-1$ boxes.

We will now argue that the set of relations obtained in this way is a complete set. 
To do this, we will be using the classification of half BPS operators given in \cite{CoRa}.
The total number of linearly independent Schur polynomials is equal to the total number
of Schur polynomials minus the number of relations between them. The number of Schur 
polynomials for a given number $n$ boxes is equal to the number of irreducible inequivalent
representations of the permutation group $S_n$. This is in turn equal to the number
of partitions of $n$. This can be computed by reading off the power of $x^n$ in the 
expansion of the product

$$ f_1(x)=\prod_{m=1}^{\infty}{1\over 1-x^m}\equiv \sum_{n=0}^\infty c_n x^n.$$

\noindent
The number of linearly independent polynomials built from $n$ $\Psi$s is equal to the number
of partitions of $n$ which do not include any 1s in the partitions. Excluding 1s accounts for
the fact that products including any factors of $\Tr (\Psi)$ vanish. This can be computed 
by reading off the power of $x^n$ in the expansion of the product

$$ f_2(x)=\prod_{m=2}^{\infty}{1\over 1-x^m}\equiv \sum_{n=0}^\infty d_n x^n.$$

\noindent
Noting that $(1-x)f_1 (x)=f_2(x)$, we see that

$$ d_n =c_n-c_{n-1}.$$

\noindent
Thus, the total number of relations between Schur polynomials corresponding to Young diagrams
with $n$ boxes is equal to the number of Schur polynomials with $n-1$ boxes, which is precisely
equal to the number of constraints we found above.

The local operators in ${\cal N}=4$ super Yang-Mills theory can be organized into 
irreducible representations of the $D=4$ ${\cal N}=4$ superconformal algebra. Each 
irreducible representation contains special operators of lowest scaling dimension 
related to their ${\cal R}$ charge, the so-called chiral primary operators. They 
transform in the $(0,l,0)$ representation of the $SU(4)$ ${\cal R}$ symmetry. From each 
of these $(0,l,0)$ representations we can select a unique state built from a product 
of $l$ $\Psi$s. Since the ${\cal R}$ symmetry transformation of our operator is 
independent of how we choose to contract the gauge indices, we can count the
number of irreducible representations by counting the number of ways we have of 
contracting the gauge indices on $l$ $\Psi$s. This is of course equal to $d_l$.

\subsection{A Linearly Independent Basis of $SU(N)$ BPS Operators}

In this section we will use polynomials with a different normalization

$$\tilde{\chi}_R(\Psi )={d_R\over D_R n_R!}\chi_R (\Psi ).$$

\noindent
The advantage of this normalization is that the action of $D$ is now

$$ D\tilde{\chi}_R (\Phi )=\sum_{S}\tilde{\chi}_S (\Phi ),$$

\noindent
where the sum again runs over all valid Young diagrams that can be obtained
from $R$ by removing a single box. Notice that $D$ maps the space
of Schur polynomials associated with Young diagrams with $n$ boxes (of dimension
$c_n$) to the space of Schur polynomials associated with Young diagrams with 
$n-1$ boxes (of dimension $c_{n-1}$). Clearly, $D$ has $c_n-c_{n-1}=d_n$ zero
eigenvectors. This is equal to the number of linearly independent Schur
polynomials built from the $\Psi$s. Thus, the basis of the null space of
$D$ is in a one-to-one correspondence with the half-BPS representations of
the super Yang-Mills theory with gauge group $SU(N)$. For any null vector
of $D$, $Df(\Phi )=0$ we have the identity

$$ f(\Psi )\to f(\Phi +c{\bf 1})=f(\Phi ).$$  

\noindent
This implies a significant simplification: for this class of operators all 
computations can be performed without making use of the ghost $c$. 

In the remainder of this section we give an algorithm which provides an explicit 
construction of a basis for the null space of $D$. Towards this end we introduce 
the notion of a fixed top block Young diagram. We call the block in the first row 
in the right most position, the {\it top block} in the Young diagram. If the top
block cannot be removed to leave a legal Young diagram, we say we have a
{\it fixed top block Young diagram}. The importance of the fixed top block Young 
diagrams is that they can be put into a one-to-one correspondence with the null
vectors of $D$. If a Young diagram is not a fixed top block diagram, it is a 
moveable top block diagram. 

To illustrate our argument we will construct a basis for the null vectors of $D$ of 
${\cal R}$ charge $n$. The full set of relations between the Schur polynomials 
$\chi_R(\Psi )$ associated with Young diagrams of $n$ boxes can be used to eliminate
all of the moveable top block Young diagrams of $n$ boxes. Each relation is obtained
by taking the product of the fundamental representation ($\yng(1)$) with a representation
corresponding to a Young diagram with $n-1$ boxes. Use each such relation to eliminate the
moveable top block Young diagram obtained by adding $\yng(1)$ to the Young diagram
with $n-1$ boxes, so that the added box is in the top block position. Note that every 
moveable top block Young diagram can be obtained by adding $\yng(1)$ to a Young diagram
with $n-1$ boxes, so that the added box is in the top block position. Thus, this 
eliminates all of the moveable top block Young diagrams. 

Our algorithm constructs a null vector of $D$ from each fixed top block diagram by
using an operation we call the {\it reduction} of the Schur polynomial. The reduction 
of a Schur polynomial is obtained by taking minus one times the sum of Schur Polynomials
corresponding to diagrams that can be obtained by all moves which move a box into the first
row, such that after the move we obtain a valid Young diagram. As an example, the reduction of
$\chi_{\yng(3,2,1)}(\Psi )$ is given by

$$ -\chi_{\yng(4,1,1)}(\Psi)-\chi_{\yng(4,2)}(\Psi).$$

\noindent
As another example, the reduction of $\chi_{\yng(4)}(\Psi )$ is zero. We can obtain a higher
order reduction by reducing a reduction of the Schur polynomial. For a Schur polynomial with
$n$ boxes, we reduce at most $n$ times. We can now state our algorithm: by starting with a 
Schur polynomial corresponding to a fixed top block Young diagram and adding all possible
reductions, we obtain a null vector of $D$.

We end this section with a construction of the null vectors of $D$ of ${\cal R}$ charge 4.
There are two fixed top block Young diagrams ($\yng(1,1,1,1)$ and $\yng(2,2)$). Applying
our algorithm we obtain two null vectors

$$\chi_1(\Psi)=\chi_{\yng(1,1,1,1)}(\Psi )-\chi_{\yng(2,1,1)}(\Psi )
+\chi_{\yng(3,1)}(\Psi )-\chi_{\yng(4)}(\Psi )$$

\noindent
and

$$\chi_2(\Psi )=\chi_{\yng(2,2)}(\Psi )-\chi_{\yng(3,1)}(\Psi )+\chi_{\yng(4)}(\Psi ).$$

\noindent
It is easy to check that

$$ \langle\chi_1(\Psi)\chi_2(\Psi^*)\rangle =\langle\chi_1(\Phi)\chi_2(\Phi^*)\rangle
\ne 0.$$

\noindent
Thus, the basis constructed using our algorithm is not an orthogonal basis. To see that
it is indeed linearly independent is easy: the question of linear dependence of these
operators can be settled by computing two point functions. To compute these, we
can replace all $\Psi$s by $\Phi$s, thanks to the fact that all of our operators
are null vectors of $D$. Upon replacing all $\Psi$s by $\Phi$s the linear independence
of the basis follows because Schur polynomials of the $\Phi$s corresponding to distinct
Young diagrams are linearly independent and each fixed top block Young diagram appears
in a unique operator belonging to our basis.

If the mixing between these operators was suppressed as $N\to\infty$, they would still
have formed natural candidates for particle states in the classical limit of the dual
quantum gravity. It is however easy to check that this mixing
is not suppressed as $N\to\infty$.

\subsection{Correlators of traces in the $SU(N)$ theory}

Given the correlators of Schur polynomials, following \cite{CoRa} we can compute
correlators of the form

$$\langle \Tr (\sigma_1 \Psi)\Tr (\sigma_2 \Psi )\rangle ,$$

\noindent
with $\sigma_1,\sigma_2\in S_n$ for any $n$. For $n$ of order $\sim 1$ these correspond to
Kaluza-Klein modes. Upon using the orthogonality of the characters of $S_n$ 

$$\sum_R\chi_R (\tau )\chi_R(\sigma )=\sum_{\gamma}\delta 
(\sigma^{-1}\gamma\tau\gamma^{-1}),$$

\noindent
we have 

$$\langle \Tr (\sigma_1 \Psi)\Tr (\sigma_2 \Psi )\rangle =
\sum_{R,S}\chi_R (\sigma_1)\chi_S (\sigma_2)\langle
\chi_R (\Psi )\chi_S (\Psi^* )\rangle .$$

\noindent
All sums in this subsection run over the representations of $S_n$.

\section{Giant Gravitons}

In this section, we use the technology developed above to study giant gravitons. We are
interested in studying operators in the $SU(N)$ super Yang-Mills which are
dual to giant graviton states in the $AdS_5\times S^5$ string theory. By studying the
contribution of bulk and boundary degrees of freedom, we also discuss how effectively 
giant gravitons probe the geometry of the dual gravitational theory.

Recent work has suggested that a matrix theory description for DLCQ string theory in
the $AdS_5\times S^5$ background can be constructed using gravitons with unit angular 
momentum ("tiny gravitons")\cite{S1}. For a study of further properties of giant gravitons 
using the dual field theory see\cite{Silva}. Finally, for a study of giant gravitons
in the pp-wave background and in background $B$ fields, see\cite{S2}.

\subsection{Large $N$}

When the Higgs fields can be simultaneously diagonalized, their eigenvalues can be
interpreted as transverse coordinates for the branes. The fact that $\Tr (\Psi )=0$
implies that the center of mass of all of the branes is fixed.  In view of this, 
the linearly independent basis constructed above is natural - it consists 
of operators that are annihilated by $D$, which is essentially the center of mass 
momentum. Strictly speaking, this constraint on the center of mass motion 
implies that we cannot have single eigenvalue dynamics\footnote{This is a bit dramatic.
As long as the center of mass motion factorizes we could easily remove it. This
requirement selects a privileged (set of) bases of states of the unconstrained theory.}. 
What then is a natural candidate for the field theory
dual to a giant graviton (D-brane) state? In the large $N$ limit,
there is a natural answer to this question. We could imagine changing a single eigenvalue 
by an amount, say $\Delta$, and then compensating by changing each of the remaining $N-1$ 
eigenvalues by an amount $-{\Delta\over N-1}$. This preserves $\Tr (\Psi)=0$.
In a systematic large $N$ expansion, at leading order, we can ignore the 
$-{\Delta\over N-1}$ effect and hence in this limit, we recover single 
eigenvalue dynamics. Thus, we again expect the Schur polynomials to be the correct
operators dual to the giant graviton (D-brane) states.

As a test of this idea, we should recover orthogonality between Schur 
polynomials corresponding to totally symmetric representations (dual to AdS giants) 
and Schur polynomials corresponding to totally antisymmetric representations (dual to
sphere giants) in this limit. Using the techniques developed in section 3, we explicitly 
check this. It is important to stress that it is not obvious that these two Schur polynomials 
becomes approximately orthogonal in the large $N$ limit. Indeed, it is easy to show 
that ratios like

$$ {\langle \chi_{\yng(5,1)}(\Psi )\chi_{\yng(6)}(\Psi^*)\rangle \over
\langle \chi_{\yng(6)}(\Psi )\chi_{\yng(6)}(\Psi^*)\rangle},$$

\noindent
and

$${\langle \left(\chi_{\yng(4,1,1)}(\Psi )+\chi_{\yng(4,2)}(\Psi )\right)
\chi_{\yng(6)}(\Psi^*)\rangle\over\langle \chi_{\yng(6)}(\Psi )\chi_{\yng(6)}(\Psi^*)\rangle } .$$

\noindent
are of order $\sim 1$ {\it independent} of $N$. Similar ratios can be written involving Schur 
polynomials with an arbitrarily large number of boxes in the Young diagram, and these inner
products remain of order $\sim 1$ independent of the number of boxes. We now turn 
to checking the expected orthogonality.

Concretely, we would like to compute

$${{\cal O}\over {\cal N}_S{\cal N}_A},$$

\noindent
where, for the case where our states correspond to representations with 5 boxes we
would have

$$ {\cal O}=\langle \chi_{\yng(5)}(\Psi )\chi_{\yng(1,1,1,1,1)}(\Psi^*)\rangle,$$

$${\cal N}_A= \sqrt{\langle \chi_{\yng(1,1,1,1,1)}(\Psi )\chi_{\yng(1,1,1,1,1)}(\Psi^*)\rangle},$$

$${\cal N}_S=\sqrt{\langle \chi_{\yng(5)}(\Psi )\chi_{\yng(5)}(\Psi^*)\rangle}.$$

\noindent
We are interested in computing these quantities as a function of the number of boxes, which
we denote by $n$. The computation of ${\cal O}$ is much simpler than the computation of
${\cal N}_S$ or ${\cal N}_A$, because for ${\cal O}$, only the $(n-1)$th and $n$th terms in 
the Taylor expansion contribute. For both ${\cal N}_S$ and ${\cal N}_A$ contributions need
to be summed from all $n+1$ terms in the Taylor expansion. For ${\cal O}$ we obtain

\begin{eqnarray}
{\cal O}&=&\prod_{i=1}^{n-1}(N^2-i^2)\left[
{(-1)^{n-1}\over (n-1)!N^{n-1}}
\langle\chi_{\yng(1)}(\Psi)\chi_{\yng(1)}(\Psi^*)\rangle
+{(-1)^n N^2\over N^{n}n!}\right]
\nonumber\\
&=&{(-1)^{n-1}(n-1)\over n!N^{n-2}}
\prod_{i=1}^{n-1}(N^2-i^2)
\nonumber\\
&=&{(-1)^{n-1}(n-1)\over n!N^{n-1}}
{(N+n-1)!\over (N-n)!}.
\nonumber
\end{eqnarray}

\noindent
In a similar way

$${\cal N}_S^2={(N+n-1)!\over (N-1)!}\sum_{i=0}^{n}{(-1)^i\over N^i}
{(N+n-1)!\over i!(N+n-i-1)!},$$

$${\cal N}_A^2={N!\over (N-n)!}\sum_{i=0}^{n}{(-1)^i\over N^i}
{(N+i-n)!\over i!(N-n)!}.$$

\noindent
Putting these results together, the quantity we are interested in is

$$ {{\cal O}\over {\cal N}_S{\cal N}_A}=
{(-1)^{n-1}(n-1)\over n! N^{n-1}}
\left[
\sum_{i,j=0}^n {(-1)^{i+j}(N-n+j)!\over N^{i+j-1}i!j! (N+n-i-1)!}
\right]^{-{1\over 2}}.$$

\noindent
We do not have a closed form answer for this quantity. It is however
easy to compute this number numerically. The result is shown in the figure below.
Thus, the mixing between these states is not suppressed for small $n$, but rapidly 
goes to zero as $n$ is increased. Note that for any value of $N$,
$\left|{{\cal O}\over {\cal N}_S{\cal N}_A}\right|=1$ for $n=2,3$. This is the 
maximum value that the overlap can attain, indicating that these two states 
are identical up to a sign.

The results have a clear physical interpretation. For small values of $n$ the operators
that we are studying are dual to states in the supergravity with a small mass. The wave 
functions of these states will not be well localized - light objects are described by
wave functions with large position fluctuations. For these states we thus expect significant
mixing of the bulk and boundary degrees of freedom. Consequently the difference between
operators in the $SU(N)$ theory (which does not have the boundary degrees of freedom) and 
the $U(N)$ theory (which does) is large and we have no right to expect that we will inherit 
the orthogonality between the corresponding operators of the $U(N)$ theory, even in the 
large $N$ limit.

{\vskip -2.5 cm}
\begin{center}
\begin{figure}[h]{\psfig{file=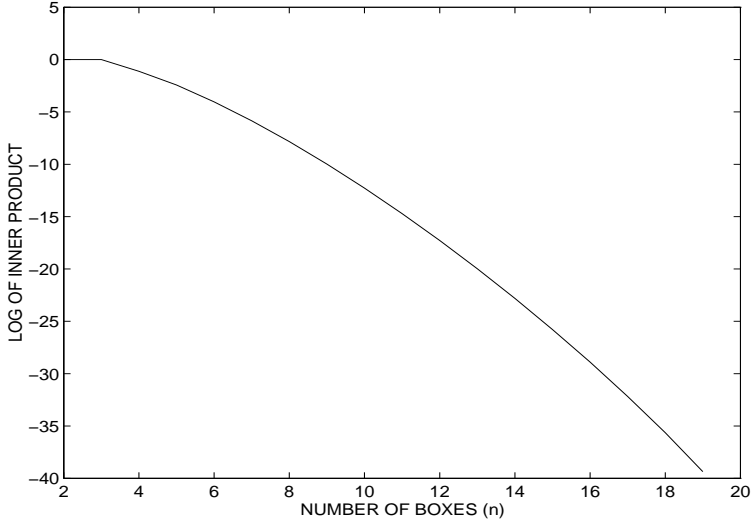,width=10cm,height=7.0cm}
 \caption{A plot showing $\log\left|{{\cal O}\over {\cal N}_S{\cal N}_A}\right|$ as a
function of the number of boxes $n$, for $N=20$.}}
\end{figure}
\end{center} 
{\vskip -1.1cm}

In the case when $n$ is of order $\sim N$, our operators are dual to heavy objects whose 
wave functions will be well localized in the bulk of the AdS space. For these states we 
expect much less mixing of the boundary and bulk degrees of freedom. Consequently, we 
don't expect that the boundary degrees of freedom make a large contribution to the state, 
and the difference between the operators in the $SU(N)$ gauge theory and the $U(N)$ gauge 
theory is not great. In this case we {\it do} expect to inherit the orthogonality of the
$U(N)$ operators. 

\subsection{Finite $N$}

At finite $N$, there is not even an approximate notion of single eigenvalue dynamics.
In this case, perhaps the simplest way forward is to
apply the Gram-Schmidt algorithm to the linearly independent operators selected in 
section 3.4 to obtain a new set of operators which do diagonalize the inner product. This
process can be used to produce many different sets of operators diagonalizing the two point
function. Which set is the most useful for the gauge theory/gravity dictionary? We do not
have a satisfactory answer to this question. 

Although the Gram-Schmidt algorithm can be used to diagonalize the two point function, this 
does not give a practical solution when the number of boxes in the Young diagrams becomes 
large. Is there a more elegant way to perform the orthogonalization? 

\noindent
The orthogonality of 
Schur polynomials in the $\Phi$ field was proved using the Frobenius-Schur duality between 
the symmetric and unitary groups\cite{JeRa}. It is natural to ask if there is a 
generalization of these results for the Schur polynomials in the $\Psi$ field. There is 
a simple possibility that can be considered. Irreducible representations of the unitary 
groups are obtained by taking contractions of objects transforming in the fundamental
representation with tensors that have a definite symmetry under interchange of indices.
Irreducible representations of the orthogonal groups are obtained by taking contractions
with {\it traceless} tensors that have a specific symmetry under interchange of indices. 
One might guess that this tracelessness constraint is exactly what is needed to solve the
problem studied here.

The relevance of the permutation group for Frobenius-Schur duality comes from the fact that
the permutation group and unitary groups are centralizers of each other. The centralizers 
of the orthogonal group are the Brauer algebras. Thus, one may have suspected that the 
polynomials (in $\Psi$) built by replacing the characters of the permutation group in the 
Schur polynomials by the characters of the Brauer algebra\footnote{A useful reference 
for characters of the Brauer algebra is\cite{BA}.} would have diagonal two point functions.
We have explicitly checked that this guess is not correct.

\subsection{Giant Gravitons as probes of the Dual Geometry}

Natural probes of the dual geometry are heavy objects which can be treated
semiclassically. The mass of the object is set by the scale on which the geometry
is to be probed. If a scale $\gamma$ is to be resolved, the fluctuations in the
position of the probe must be less than $\gamma$, forcing the mass of the probe to 
be larger than $\gamma^{-1}$. Of course, there is a smallest scale which can
be probed. At this smallest scale the probe starts to noticeably deform the 
background metric, that is, its gravitational radius is no longer smaller than $\gamma$. 
This smallest scale is the Planck scale.

When the gauge theory is at strong coupling, the radius of AdS is much larger than 
the string scale. In this limit and at geometric distances which are larger than the 
string scale, we'd expect that giant gravitons are good probes of the geometry.

For the gravity dual to the $U(N)$ gauge theory, since there are both bulk and boundary
degrees of freedom, it is not good enough to simply require a heavy probe.
Indeed, we could imagine a composite object composed of a heavy excitation of boundary 
modes and a light excitation of the bulk gravity. Even if the mass of the composite
is larger that $\gamma^{-1}$, this probe might not resolve features in the bulk of
order $\gamma$. In this context, it is appropriate to ask if we can separate the
bulk and boundary contributions to operators in the dual gauge theory.

\subsection{Disentangling bulk and boundary degrees of freedom}

Up to this point, we have used a ghost field $c$ to subtract the mode corresponding
to the trace of $\Phi$. It is also possible to express $\Phi$ in terms of $\Psi$ by
adding the mode corresponding to the trace. In terms of the $b$ field

$$\langle bb\rangle=\langle b^* b^*\rangle,\qquad \langle bb^*\rangle ={1\over N},$$

\noindent
we have the identity

$$\langle F(\Phi_{ij},\Phi_{kl}^*)\rangle = 
\langle F(\Psi_{ij} +b\delta_{ij},\Phi_{kl}^* +b^*\delta_{kl})\rangle ,$$

\noindent
for an arbitrary function $F(\cdot,\cdot )$. In contrast to our previous computations,
where the ghost had no physical meaning, $b$ and $\Psi$ do have clear physical interpretations.
$b$ describes the overall center of mass motion of all of the branes; in the gravity dual it
is localized on the boundary of $AdS_5\times S^5$. $\Psi$ is describing the bulk string theory.
The technology we have developed above can now be used to disentangle bulk and boundary degrees
of freedom in the dual super Yang-Mills theory. For concreteness, consider the operator
dual to a sphere giant, that is, a Schur polynomial corresponding to an antisymmetric
representation with $n$ boxes and $n\sim N$. This operator is dual to a heavy state and so
should provide a good (well localized) probe of the geometry of the gravitational theory.  

Let ${\cal R}_n$ denote the Young diagram with a single column and $n$ rows. If $n=0$,
${\cal R}_n=1$. Then, we have

$$\chi_{{\cal R}_n}(\Phi )\to \chi_{{\cal R}_n}(\Psi +b{\bf 1})=
\sum_{i=0}^n {b^i\over i!}\prod_{k=1}^{i} (N-n+k)
\chi_{{\cal R}_{n-i}}(\Psi ).$$

\noindent
Terms in the above sum corresponding to low values of $i$ are the product of a Schur
polynomial corresponding to a representation with a large number of boxes and a boundary
state with a small ${\cal R}$ charge. Thus, these terms correspond to objects in the bulk
gravity which are heavy and hence well localized times light boundary excitations which will
not be well localized.  Terms corresponding to large values of $i$ are the product of a Schur
polynomial corresponding to a representation with a small number of boxes and a boundary
state with a large ${\cal R}$ charge. Thus, these terms correspond to objects in the bulk
gravity which are light and hence not well localized times heavy boundary excitations which 
will be well localized. To understand why, in spite of this, the Schur polynomials still
provide localized probes, recall that the two point functions of the Schur polynomials are 
not normalized. To estimate how large the different contributions to $\chi_R(\Phi)$ are, we 
should work with properly normalized operators. In terms of ${b^{\prime n}\over\sqrt{n!}}$ and
$\bar{\chi}_{{\cal R}_n}(\Psi )$, with

$$ \langle{{b}^{\prime i}\over \sqrt{i!}}\bar{\chi}_{{\cal R}_{n-i}}(\Psi )
{b^{\prime * j}\over\sqrt{j!}}\bar{\chi}_{{\cal R}_{n-j}}(\Psi^* )\rangle=\delta_{ij},$$

\noindent
we have

$$\chi_{{\cal R}_n}(\Phi )\to \chi_{{\cal R}_n}(\Psi +b{\bf 1})=
\sum_{i=0}^n\alpha_i{b^{\prime i}\over \sqrt{i!}}\bar{\chi}_{{\cal R}_{n-i}}(\Psi ),$$

\noindent
with

$$\alpha_i={(N-n+i)!\over (N-n)!}{{\cal N}_A(n-i)\over\sqrt{ N^i i!}},\qquad
{\cal N}_A^2(n)={N!\over (N-n)!}\sum_{i=0}^{n}{(-1)^i\over N^i}
{(N+i-n)!\over i!(N-n)!}.$$

{\vskip -0.6cm}
\begin{center}
\begin{figure}[h]{\psfig{file=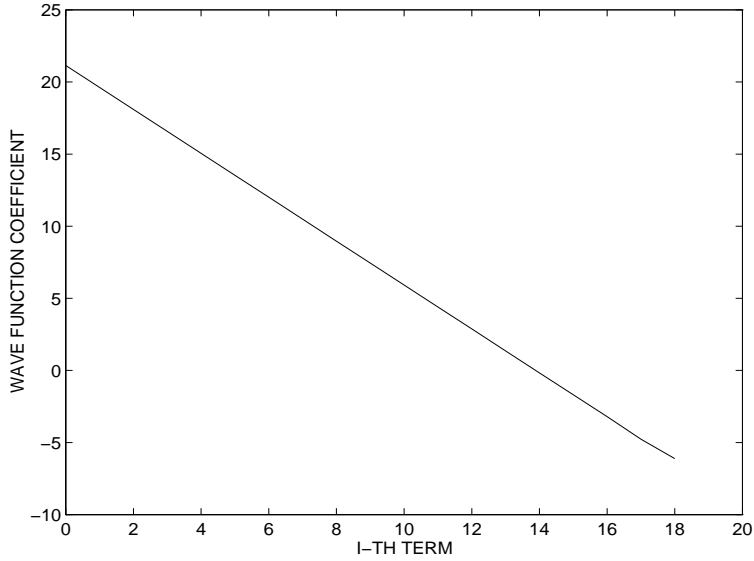,width=10cm,height=7.5cm}
 \caption{This plot shows the components in the operator dual to a sphere giant.
$\log\left|\alpha_i\right|$ is plotted as a
function of $i$, for $N=20$ and $n=20$.}}
\end{figure}
\end{center} 
{\vskip -0.6cm}

Clearly, the above expansion for the operator $\chi_{{\cal R}_n}(\Phi )$ is dominated
by terms corresponding to low values of $i$. Consequently, the contribution from the
boundary degrees of freedom to these operators is exponentially suppressed. Thus, the 
$\chi_{{\cal R}_n}(\Phi )$ do provide localized probes of the dual geometry. 

What we have found is that, in the large $N$ limit and for $n\sim N$ boxes, the
dependence on the $b$ modes is suppressed. Consequently Schur 
polynomials for the totally antisymmetric representations in $\Phi$ essentially 
coincide with Schur polynomials for the totally antisymmetric representations in $\Psi$. 
Evidently Schur polynomials in the $SU(N)$ theory corresponding to the totally 
antisymmetric representations with $n\sim N$ boxes again provide the duals to sphere 
giants.

Finally, the corresponding discussion for AdS giants is equally straightforward.
In this case we have (in what follows ${\cal S}_n$ denotes a Young diagram with a single
row of $n$ boxes)

$$\chi_{{\cal S}_n}(\Phi )\to \chi_{{\cal S}_n}(\Psi +b{\bf 1})=
\sum_{i=0}^n\alpha_i{b^{\prime i}\over \sqrt{i!}}\bar{\chi}_{{\cal S}_{n-i}}(\Psi ),$$

\noindent
with

$$\alpha_i={(N+n-1)!\over (N+n-i-1)!}{{\cal N}_S(n-i)\over\sqrt{N^i i!}},\qquad
{\cal N}_S^2(n)={(N+n-1)!\over (N-1)!}\sum_{i=0}^{n}{(-1)^i\over N^i}
{(N+n-1)!\over i!(N+n-i-1)!}.$$

{\vskip -0.5cm}
\begin{center}
\begin{figure}[h]{\psfig{file=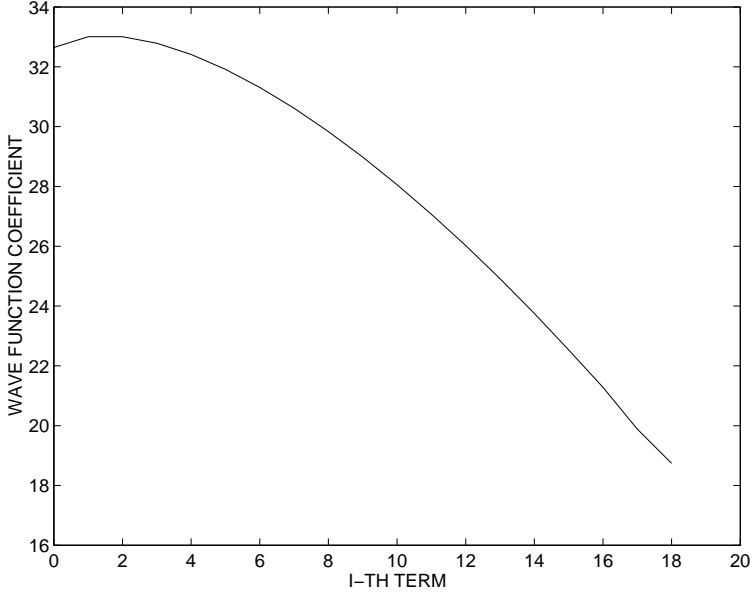,width=10cm,height=8cm}
 \caption{This plot shows the components in the operator dual to an AdS giant.
$\log\left|\alpha_i\right|$ is plotted as a
function of $i$, for $N=20$ and $n=20$.}}
\end{figure}
\end{center}

It is interesting to note that the coefficients $\alpha_i$ reach a maximum at $i=1,2$.
For these terms there is a very small contribution coming from the boundary degrees 
of freedom. Clearly however, the boundary contribution is still significantly suppressed.

\section{Summary}

The results of \cite{JeRa} provide a set of operators, the Schur polynomials, which 
have diagonal two point functions and are in one-to-one correspondence with the 
half-BPS representations of the $U(N)$ gauge theory at zero coupling and finite $N$. 
Obtaining the corresponding result for the $SU(N)$ gauge theory was one of the goals
of this article.

We have developed the necessary technology to study Schur polynomials in the $SU(N)$ 
gauge theory. In addition, we have obtained a complete set of relations between the
Schur polynomials and have provided an algorithm which selects a unique linearly
independent basis for the half-BPS representations. Using the Gram-Schmidt algorithm
we can obtain a new set of operators with diagonal two point functions. Although this
answer is not very explicit, it does provide the generalization we were after. 

We have argued that at large $N$ Schur polynomials corresponding to either 
totally symmetric or totally antiymmetric representations with a large number 
of boxes, are approximately dual to giant graviton states. Using the techniques
developed in section 3, we were able to study the two point function of these two 
operators and verify that it goes to zero exponentially fast as the number of 
boxes is increased.

Further, our technology when applied to the $U(N)$ gauge theory allows for a
separation of the bulk and boundary degrees of freedom. We have verified that 
for both sphere giants and AdS giants, the boundary contribution is small enough for
the Schur polynomials of the $SU(N)$ gauge theory to remain good candidate duals to giant
gravitons. 

$$ $$

\noindent
{\it Acknowledgements:} 
We would like to thank Eric Gimon, Antal Jevicki, Jo\~ao Rodrigues and especially
Sanjaye Ramgoolam for pleasant discussions and David Berenstein for helpful
correspondence. We would also like to thank Sanjaye for helpful comments on the 
manuscript. This work is supported by NRF grant 
number Gun 2047219.

\end{document}